\documentclass[11pt,a4paper]{article}

\usepackage[latin1]{inputenc}
\usepackage{amsmath}
\usepackage{amsfonts}
\usepackage{amssymb}
\usepackage{geometry}

\begin{document}

\title{The Cosmological Constant\\
 Emerging From Local Poincar\'e Invariance}

\author{Paul von der Heyde\thanks{Paul von der Heyde,
    J.-H.-Campe-Str.3, D-37627 Deensen, Germany, E-mail p\_vdh@gmx.de}}

\date{14 July 2005, \it{ file vdhcc.tex}}

\maketitle

\begin{abstract}
The Poincar\'{e} Gauge Theory of gravitation with a Lagrangian quadratic in the field strengths is
applied to a classical cosmological model. It predicts a constant value of the non-riemannian curvature
scalar, which acts as a cosmological constant. As the value of the scalar depends on the context, vacuum
solutions may differ from the predictions based on Einstein's constant. The corresponding deviations
from General Relativity are discussed on the basis of exact solutions for the field of a mass point.
\end{abstract}
\vspace{5mm}

The Poincar\'{e} Gauge Theory (PGT) \cite{ref1} transfers the concept of a Yang-Mills theory to gravity.
Gauge transformations are 4 spacetime translations (parallel displacements) and 6 rotations (3 spatial
rotations and 3 Lorentz boosts). Translations and rotations are independent of each other. The gauge
potentials related to translation are the tetrad $A_i{}^\alpha(x)\,,\;\alpha = \hat 0, \hat 1,\hat 2,
\hat3,\;$ i.e. orthonormal basis vectors of the tangential Minkowski space at the event $x^i,\, i =
0,1,2,3$. Tetrad indices like $\alpha$ are raised and lowered by the Minkowski metric $g_{\alpha\beta} =
\hbox{diag}(1,-1,-1,-1)$, coordinate indices like $i$ by means of the Riemannian metric $g_{ij} =
A_i{}^\alpha A_j{}^\beta g_{\alpha\beta}\,$, respectively. The potentials related to rotation form the
connection $A_{i\alpha}{}^\beta(x)\,$ of a non-Riemannian spacetime $U_4\,,$ characterized by the
antisymmetry $A_{i\alpha\beta} = A_{i[\alpha\beta]} := (A_{i\alpha\beta} - A_{i\beta\alpha})/2$.
The corresponding field strengths are the torsion tensor $F_{ij}{}^\alpha$ and the curvature tensor
$F_{ij\alpha}{}^\beta$ of the $U_4\,$:
\begin{equation}\label{eq:1}
  F_{ij}{}^\alpha \::=\: 2\:(\,\partial_{[i}A_{j]}{}^\alpha
    + A_{[i|\gamma}{}^\alpha A_{|j]}{}^\gamma\,)\;,\qquad
  F_{ij\alpha}{}^\beta \::=\: 2\:(\,\partial_{[i}A_{j]\alpha}{}^\beta
    + A_{[i|\gamma}{}^\beta  A_{|j]\alpha}{}^\gamma\,)\;.
\end{equation}
The coupling of a special relativistic Lagrangian to gravity $L(\psi,\partial_i\psi)\Rightarrow
\mathcal L(\psi,\partial_i\psi,\mathcal A_i{}^\alpha,A_{i\alpha}{}^\beta)$ refers the matter variables
$\psi$ to the tetrad or its inverse $A^i{}_\alpha = (\partial A /\partial A_i{}^\alpha)/A\,,\;A = 
\det A_i{}^\alpha$ and replaces $\partial_{i}\,\psi$ by the covariant derivative $D_i\,\psi =
(\partial_i + A_{i\alpha}{}^\beta f_\beta{}^\alpha)\,\psi$. The canonical currents of matter, identified
by the invariance of the action under local Poincar\'e transformations, are the energy-momentum
$\mathcal J^i{}_\alpha$ as a source of torsion and the material spin $\mathcal J^i{}_{\alpha\beta}$
as a source of curvature:
\begin{equation}\label{eq:2}
  \mathcal J^i{}_\alpha := \frac{\delta\mathcal L}{\delta A_i{}^\alpha} = -\frac{\partial\,\mathcal L}
	{\partial(\partial_i\psi)}\,D_\alpha\,\psi + A^i{}_\alpha\,\mathcal L \;,\qquad
  \mathcal J^i{}_{\alpha\beta} := \frac{\delta \mathcal L}{\delta A_i{}^{\alpha\beta}} =
	-\frac{\partial\mathcal L}{\partial(\partial_i\psi\,)}\,f_{\alpha\beta}\:\psi\;.
\end{equation}
In this paper we restrict the total Lagragian $\mathcal V = \mathcal L \:+ \stackrel{t}{\mathcal L}
(A_i{}^\alpha, F_{ij}{}^{\alpha})\:+\stackrel{r}{\mathcal L} (A_i{}^{\alpha}, F_{ij}{}^{\alpha\beta})$
to be purely quadratic in the corresponding field strengths. Mixing terms of torsion and curvature or
combinations with dual tensors don't occur.  We omit the linear curvature scalar $F =
F_{\alpha\beta}{}^{\beta\alpha}$ as well, for reasons to be given later. Variation with respect to the
potentials yields the general form of the field equations
\begin{align}\label{eq:3}
 &D_j\mathcal H^{ij}{}_\alpha =
    \mathcal J^i{}_\alpha + {\stackrel{t}{\mathcal J}}{}^i{}_\alpha +
    {\stackrel{r}{\mathcal J}}{}^i{}_\alpha\:,\qquad
  D_j\mathcal H^{ij}{}_{\alpha\beta} =
    \mathcal J^i{}_{\alpha\beta} + {\stackrel{t} {\mathcal J}}{}^i{}_{\alpha\beta}\,,\\[3mm]
 &\mathcal H^{ij}{}_\alpha :=
    - 2 \frac{\partial\,\mathcal V}{\partial F_{ij}{}^\alpha} =
    \mathcal H^{[ij]}{}_\alpha \:, \qquad\;\;\;
  \mathcal H^{ij}{}_{\alpha\beta} :=
    -2\frac{\partial\,\mathcal V}{\partial F_{ij}{}^{\alpha\beta}} =
    \mathcal H ^{[ij]}{}_{\alpha\beta} = \mathcal H^{ij}{}_{[\alpha\beta]}\,,\\[2mm]
 &\stackrel{t}{\mathcal J}{}^i{}_\alpha :=
   \mathcal H^{ik}{}_\gamma  F_{\alpha k}{}^\gamma + A^i{}_\alpha \stackrel{t}{\mathcal L}\:,\qquad\;\:
 \stackrel{r}{\mathcal J}{}^i{}_\alpha := \mathcal
    H^{ik}{}_{\beta\gamma}F_{\alpha k}{}^{\beta\gamma}+ A^i{}_\alpha \stackrel{r}{\mathcal L}\:,\qquad
 \stackrel{t}{\mathcal J}{}^i{}_{\alpha\beta}:= \mathcal H^i{}_{[\beta\alpha]}\,.
\end{align}
The energy-momentum densities $\stackrel{t}{\mathcal J}$ and $\stackrel{r}{\mathcal J}$ of the tetrad
and the connection, respectively, are traceless:
\begin{equation}\label{eq:6}
 \stackrel{t}{\mathcal J}{}^\alpha{}_\alpha=
	{\stackrel{r}{\mathcal J}}{}^\alpha{}_\alpha= 0 \qquad \Longrightarrow \qquad
	A_i{}^\alpha\:D_j\mathcal H^{ij}{}_\alpha = \mathcal J^\alpha{}_\alpha\,.
\end{equation}
According to (3)$_2$, the curvature is exclusively generated by the spin density
${\stackrel{t}{\mathcal J}}{}^i{}_{\alpha\beta}$ of the tetrad and the spin density of matter.
The connection itself possesses no gauge invariant spin \cite{ref2}.\bigskip

Despite of the different structures of the PGT and General Relativity (GR), the predicted metrical
properties of spacetime coincide largely. As for the metric, the curvature scalar of a Riemann space
is formally identical with a certain combination of torsion squares in a $U_4$. In addition, as will
be shown now, the coupling constant of the gravitational spin interaction turns out to be extremely small.\bigskip

In order to demonstrate the rise \cite{ref3} of the cosmological constant $\Lambda_{\textsc{PGT}}$,
it is sufficient to consider a simple model of a universe, filled with a classical gas. Using natural
units $(\hslash=c=1),$\\we prescribe $\mathcal{J}^i{}_{\alpha\beta}=0\,,\;\mathcal{J}^i{}_\alpha = A\;
\hbox{diag}(0,0,0,M/R^3)$, with the constant mass $M$ and a distance scale $R(t)$, and the tetrad of the
flat Robertson-Walker metric in polar coordinates $(x^0,x^1,x^2,x^3) \equiv (t,r,\vartheta, \varphi)$:
\begin{equation}\label{eq:7}
  A_0{}^{\hat 0}= 1\,,   \quad
  A_1{}^{\hat 1}= R\,,   \quad
  A_2{}^{\hat 2}= R\,r\,,\quad
  A_3{}^{\hat 3}= R\,r\sin\vartheta\,.
\end{equation}
In an isotropic, homogeneous, and PT-invariant universe, the connection essentially contains only one
dynamic component $C(t)\,$, effecting a Lorentz boost in direction of the parallel displacement:
\begin{equation}\label{eq:8}
  A_{1\hat 1}{}^{\hat 0} = C\,,\;\; 
  A_{2\hat 2}{}^{\hat 0} = C\:r\,,\;\; 
  A_{3\hat 3}{}^{\hat 0} = C\:r\,\sin\,\vartheta\,,\;\; 
  A_{2\hat 1}{}^{\hat 2} = 1\,,\;\;                     
  A_{3\hat 1}{}^{\hat 3} = \sin\,\vartheta\,,\;\;  
  A_{3\hat 2}{}^{\hat 3} = \cos\,\vartheta\,.
\end{equation}
The components of the field strengths reduce to
$(\;\,\dot{R} \equiv \partial\,_t\,R\,, \;\;\dot{C} \equiv \partial\,_t\,C\;)$
\begin{align}\label{eq9}
  &F_{\hat 0 \hat 1}{}^{\hat 1} \:=\: F_{\hat 0 \hat 2}{}^{\hat 2}\:=\:
   F_{\hat 0 \hat 3}{}^{\hat 3} \,= \quad (\dot{R}-C)/R\,, \\
  &F_{\hat 0 \hat 1}{}^{\hat 0 \hat 1} = F_{\hat 0 \hat 2}{}^{\hat 0 \hat 2} =
   F_{\hat 0 \hat 3}{}^{\hat 0 \hat 3} = \quad \dot C/R\,,
  &F_{\hat 1 \hat 2}{}^{\hat 1 \hat 2} = F_{\hat 1 \hat 3}{}^{\hat 1 \hat 3} =
   F_{\hat 2 \hat 3}{}^{\hat 2 \hat 3} = \quad (\,C/R\,)\,^2\,.
\end{align}
With $F_\alpha{}^i := F_{\alpha\,\beta}{}^{\beta\,i}$ and $ F := F_\alpha{}^\alpha$,
the quadratic field Lagrangian (restricted above) reads
\begin{align}\label{eq:11}
  &\stackrel{t}{\mathcal{L}} = \frac{1}{4} A\,F_{ji}{}^{\alpha} (\;
       k_1\, F^{ij}{}_\alpha + 
    2\,k_2\, F_\alpha{}^{ij} +
    2\,k_3\, A^i{}_\alpha F^{j\beta}{}_\beta \;)\:, \\[1mm] \nonumber
  &\stackrel{r}{\mathcal{L}} = \frac{1}{4} A\, F_{ji}{}^{\alpha\beta} (\;
       k_4\, F^{ij}{}_{\alpha\beta} +  
       k_5\, F_{\alpha\beta}{}^{ij} + 
    4\,k_6\, F^i{}_{\alpha\beta}\,^j + 
    4\,k_7\, A^i{}_\alpha F_\beta\,^j +
    4\,k_8\, A^i{}_\alpha F^j{}_\beta + 
    4\,k_9\, A^i{}_\alpha A^j{}_\beta \, F \;)\:.
\end{align}

\pagebreak

\noindent Because of the symmetry of the model, the nine constants $k_n$ join to form only two effective
constants $kt := (\,-k_1+ k_2 + 3\,k_3\,)/2\,,\; kr := k_4 + k_5 - 2\,k_6 - 4\,k_7 - 4\,k_8 - 12\,k_9\:$
entering the field equation (3)$_1$ :
\begin{align}\label{eq:12}
&6\,kt\:R\,^3\,(\,\ddot R-\dot C\;)\quad+\;\;6\,kr\:(\,R\,^2\;{\dot C}\,^2-C\,^4\,)\;\;=\;-\,M\,R\,,\\
&6\,kt\:R\,^2\,(\,{\dot R}\,^2-C\,^2\,)\;-\;6\,kr\:(\,R\,^2\;{\dot C}\,^2-C\,^4\,)\;\: =\;\;2\,M\,R\,.
\end{align}
Equation (3)$_2$, due to the lack of material spin not independent of (3)$_1$, results in the condition
\begin{equation}\label{eq:14} 
kr\,(\,R\,^2\,\ddot C + R\,\dot R\,\dot C - C\,^3\,) \;=\; kt\,R\,^2\,(\,\dot R - C\,)\,.
\end{equation}
This becomes more transparent by the substitution $\dot C=-C\,^2/R-(F\,R)/6$ gained from (10):
\begin{equation}\label{eq:15} 
kr\,R\,\dot F \;=\; 2\,(\,kr\:F + 3\,kt\,)\,(\,C - \dot R\,)\,.
\end{equation}
Here, the only surviving component of the torsion $(\,C-\dot R\,)$ must not vanish, because otherwise
$\mathcal H\,^{ij}\,_\alpha$ and, according to (6), $\,\mathcal J\,^\alpha\,_\alpha \sim M\,$
would vanish, too. The obvious solution
\begin{equation}\label{eq:16}
F\;=\;-\,\begin{Large}\frac{3\,kt}{kr}\end{Large}\,.
\end{equation}
seems to be unique, if a non-static universe and $M \neq 0$ is required \cite{ref4}.
Insertion in (12,13) yields\\
\begin{equation}\label{eq:17}
  \ddot R \;=\;
  \begin{Large}\frac{kt}{4\,kr}\begin{normalsize}R -\end{normalsize}\frac{M}{6\,kt\,R\,^2}\end{Large}
  \quad,\qquad
 {\dot R}\,^2 \;=\;
  \begin{Large}\frac{kt}{4\,kr}\begin{normalsize}R\,^2+\end{normalsize}\frac{M}{3\,kt\,R}\end{Large}\;.
\end{equation}\\
A comparison with the formal identical relations well-known from GR fixes the constants: 
\begin{equation}\label{eq:18}
kt\,=\,\begin{Large}\frac{1}{l^2}\end{Large}\;\;,\qquad
kr\,=\,\begin{Large}\frac{1}{k^2}\end{Large}\;\;,\qquad
\Lambda_{\textsc{PGT}}\,=\,\begin{Large}\frac{3\,k^2}{4\:l^2}\end{Large}\;.
\end{equation}
Here  $\,l\approx8,1*10\,^{-35}\,$m$\:$ is the Planck length, and we get $k\approx1*10\,^{-60}\,$ as a
rough value \cite{ref5} of a new coupling constant of gravitational spin interaction.\bigskip

In our simple model, the cosmological constant exclusively depends on the fundamental constants of the
theory. This result, however, does not depend on the simplicity of the model. The use of a non-flat
geometry or a more elaborated description of matter alters the equations (17) in a way known from GR,
but it does not touch the structure and solution of equation (15) as long as the spin of matter can be
neglected. We would like to stress, that in the PGT with quadratic Lagrangian, 'normal' gravity is
carried by the torsion squares, which supply the coupling $\,l^2 M$ in (17), and not by the curvature.
Adding a linear term $k_0\,A\,F$ to the Lagragian (11)$_2$ would merely shift $\Lambda_{\textsc{PGT}}$
to the value $\,3\,k^2\,(\,2\:k_0\:l^2-1\,)^2/\,(4\,l^2)$. Provided that $\Lambda_{\textsc{PGT}}\neq
0\:$, this term is ruled out, however, because only $k_0 = 0$ guarantees vanishing energy-momenta of
the gravitational fields in case of $\:t \rightarrow \infty\,$.\bigskip
 
$\Lambda_{\textsc{PGT}}\,$ differs from $\,\Lambda_{\textsc{GR}}\,$ in obvious aspects. First, the
constancy of the scalar $\,F(x)\,$ in (15) depends on the context and can only be expected in case of
high symmetry and vanishing spin. Second, in a certain sense, $\,\Lambda_{\textsc{PGT}}\,$ is induced by
matter. In the (theoretical) limit $M\rightarrow 0,\;\Lambda_{\textsc{GR}}\,$ keeps driving the
de Sitter universe apart according to $R \sim e\,^{\sqrt{\Lambda/3}\;t}$. Contrary to that,
$\,\Lambda_{\textsc{PGT}}\,$ can be switched off in that case, and with $\dot R = C = F = 0$ in (15) we
get the Minkowski space as the adequate cosmological vacuum solution.

\pagebreak

An analogous but less theoretical example is the vacuum field of an isolated mass point and the related
question, to what extent subcosmic systems, say our planetary system, are influenced by the cosmic
expansion. The embedding of a static and isotropic local metric into an expanding universe in the sense
of an Einstein-Straus model \cite{ref6} does not work in case of $\,\Lambda_{\textsc{GR}}\neq 0\,$,
as in GR such local metric is no longer available. The well-known modified Schwarzschild solution with
$\:g_{00}=1-2\,m/r-(\,\Lambda_{\textsc{GR}}/3\,)\, r^2\;$ is static, but not isotropic, and the
last term merely hides the cosmic expansion as a repulsive force. Transformed to isotropic coordinates,
however, the dynamical form of this solution
\begin{eqnarray}\label{eq:19}
&A_0{}^{\hat 0} \,=\, Q\;,\quad
A_1{}^{\hat 1} \,=\, P\,R\;,\quad
A_2{}^{\hat 2} \,=\, P\,R\,r\;,\quad
A_3{}^{\hat 3} \,=\, P\,R\,r\,\sin\vartheta\;, \\[4pt] \nonumber
&Q(r,t)\,:=\,\left(\,1-\frac{m}{2\,R\:r}\,\right)/\left(\,1+\frac{m}{2\,R\:r}\,\right)\;,\quad
P(r,t)\,:=\,\left(\,1+\frac{m}{2\,R\:r}\,\right)^2\;,\quad 
R(t)\,:=\,R_o\,e^{\sqrt{\Lambda/3}\;t}\;.
\end{eqnarray}
clearly asserts an unreduced expansion. According to (19), the orbits of the planets would have been
enlarged by roughly 45\% since the formation of the solar system \cite{ref7}, and it is hard to see,
how the impact on the climatic and biological evolution can be avoided within the framework of GR.
This problem seems not to carry such a great weight in the PGT, as far as a vacuum solution is
considered as a first approximation. Here torsion is allowed to vanish, and the isotropic solution
describing the vacuum field of a point mass is the \textit{static} Schwarzschild metric (19) with
$R=R_o$ and the connection calculated from $F_{ij}{}^\alpha = 0$.\bigskip

A peculiarity of the PGT should be mentioned here. Due to the independence of the connection, the metric
doesn't definitely fix the solution. For example, the special Lagrangian (11) with $k_1=k_3=1/\,l^2,\:  
k_4=1/\,k^2,\:k_n=0\,$ otherwise \cite{ref8}, admits three solutions with nonvanishing torsion
\cite{ref9}, which coincide in the metric (19) with $R(t)=R_o\,e^{\,\epsilon H_o\,t}$ :
\begin{align}\label{eq:20}
 &F_{\hat 1 \hat 0}{}^{\hat 0}  = \frac{2\,m\,Q}{(\,PR\,r\,)\,^2\,Q_+\,Q_-}\;,\hspace{90pt}
  F_{\hat 0 \hat 1}{}^{\hat 1}  =  \frac{2\,\epsilon\,m\,H_o}{PR\,r\;Q_+\,Q_-}\;, \\[5pt]
 &F_{\hat 1 \hat 0}{}^{\hat 0}  = - F_{\hat 1 \hat 2}{}^{\hat 2} = - F_{\hat 1 \hat 3}{}^{\hat 3}
								=  \frac{m}{(\,PR\,r\,)^2\,Q_-}\;,\qquad 
  F_{\hat 0 \hat 1}{}^{\hat 1}  =  - F_{\hat 0 \hat 2}{}^{\hat 2} = - F_{\hat 0 \hat 3}{}^{\hat 3}
                                =  \frac{\epsilon \,m}{(\,PR\,r\,)^2\,Q_-}\;, \\[5pt] 
 &F_{\hat 1 \hat 0}{}^{\hat 0}  = - F_{\hat 1 \hat 2}{}^{\hat 2} = - F_{\hat 1 \hat 3}{}^{\hat 3}
                                =  \frac{m}{(\,PR\,r\,)^2\,Q_+}\;,\qquad
  F_{\hat 0 \hat 1}{}^{\hat 1}  = - F_{\hat 0 \hat 2}{}^{\hat 2} = -  F_{\hat 0 \hat 3}{}^{\hat 3}
                                =  \frac{-\,\epsilon\,m}{(\,PR\,r\,)^2\,Q_+}\;,\\[7pt]\nonumber
 &\epsilon\,:=\,\pm 1\,,\quad H_o\,:=\,k/(2\;l)=\sqrt{\Lambda_{\textsc{PGT}}/3}\,,\qquad 
  Q_\pm\,:=\,Q \pm H_o\,PR\,r\;.
\end{align}
The meaning of these solutions is unclear. Presumably they don't describe a persistent and regular
field of a macroscopic body. Nevertheless, in our opinion, the occurrence of different states of the
gravitational field associated with one and the same source is an attractive feature of the PGT.
It indicates the existence of three different energy states of a point mass, corresponding to
the torsion-free solution, solution (20), and (21/22). The hypothesis of a relation to the mass
spectrum of elementary particles, however, remains an unproved speculation within the framework
of a classical field theory.\\

\centerline{-----------} \vspace{-10pt}
\centerline{--------}\vspace{30pt}

\noindent Many thanks to F.W.~Hehl, Cologne, for his encouragement and helpful advice. 
 
\pagebreak

\vspace{20pt}

\centerline{-----------} \vspace{-10pt}
\centerline{--------}


\begin{thebibliography}{99}
\bibitem{ref2} The same is true for gauge potentials of nongravitational interactions,
   which exclusively couple to the tetrad via $g_{ij}.$
\bibitem{ref4} We expect the stability of (16) to be limited. A sound discussion has to consider
   realistic initial conditions and the influence of material spin in the early stage of the universe.
\bibitem{ref7} This estimate simply uses the integral of (17.2) $(R/R_o)^3 = \sinh^2[(3k/4 l) t]$
   and not an approximate solution of the equations of planetary motion based on (19).
   We expect the correct result to be of the same order of magnitude.
\end{thebibliography}
\end{document}